\documentclass[preprint]{elsarticle}

\usepackage{lineno,hyperref}
\modulolinenumbers[5]

\usepackage{graphicx}
\usepackage{color}
\usepackage{multirow}
\usepackage{mathtools}
\usepackage[subfigure]{tocloft}
\usepackage[utf8]{inputenc}
\usepackage[T1]{fontenc}
\usepackage{mathptmx}
\usepackage{amssymb}
\usepackage{amsfonts}
\usepackage{array}
\usepackage{exscale}
\usepackage{physics}
\usepackage[version=4]{mhchem}
\usepackage{xfrac}
\usepackage{booktabs}
\allowdisplaybreaks
\usepackage{float}
\usepackage{subcaption}
\newcommand{\rf}[1]{\ref{#1}}
\newcommand{\eq}[1]{Eq.~(\ref{#1})}

\newcommand{\vc}[1]{\mathbf{#1}}
\newcommand{\scmag}[2]{\abs{\vc{#1}_{#2}}}

\renewcommand{\varepsilon}{\text{\usefont{OML}{cmr}{m}{n}\symbol{15}}}
\newcommand*\xbar[1]{%
      \hbox{%
        \vbox{%
          \hrule height 0.5pt % The actual bar
          \kern0.2ex%         % Distance between bar and symbol
          \hbox{%
            \kern-0.1em%      % Shortening on the left side
            \ensuremath{#1}%
            \kern-0.1em%      % Shortening on the right side
          }%
        }%
      }%
    }
\def\<{\langle}
\def\>{\rangle}
\def\e{{\, e}\,}

\def\({\Big{(}}
\def\){\Big{)}}
\def\[{\Big{[}}
\def\]{\Big{]}}
\def\nn{\nonumber}
\def\EL{\mathcal{E}_L}
\def\ER{\mathcal{E}_R}
\def\ga{\gamma^{\alpha}}

\def\gf{\gamma^5}

\def\elL{\bar{e}_L}
\def\elR{\bar{e}_R}

\def\NDBD{0\nu\beta\beta}

\def\TNDBD{2\nu\beta\beta}

\def\denstat{\frac{\dd[3]{\vc{p}}_1}{(2\pi)^32E_1}\frac{\dd[3]{\vc{p}}_2}{(2\pi)^32E_2}}
\def\Hbeta{\mathcal{H}_I^{\beta}}
\def\light{\frac{1}{m_e}\sum_k^{\text{light}}(U_{ek})^2m_k\xi_k}
\def\heavy{\lambda^2\sum_k^{\text{heavy}}(V^*_{ek})^2\frac{m_p\Xi_k}{M_k}}
\def\fermi{\bigg(\frac{G_F\,\text{cos}\,\theta_c}{\sqrt{2}}\bigg)^2}
\def\ferm{\bigg(\frac{G_F\,\text{cos}\,\theta_c}{\sqrt{2}}\bigg)}

\def\gAA{-\frac{g_A^2m_e^2}{2\pi}}

\def\EL{\mathcal{E}_L}
\def\ER{\mathcal{E}_R}

\def\gOnu{\frac{(G_F\,\cos\,\theta_c)^4m_e^9}{(2\pi)^5\ln 2}}
\begin{document}
\title{Interference Between Light and Heavy Neutrinos for $\NDBD$ Decay in the Left-Right Symmetric Model}
\author{Fahim Ahmed}
\ead{ahmed1f@cmich.edu}
\author{Andrei Neacsu}
\author{Mihai Horoi}
\address{Department of Physics, Central Michigan University, Mount Pleasant, Michigan 48858, USA}
\begin{abstract}
Neutrinoless double-beta decay is proposed as an important low energy phenomenon that could test beyond the Standard Model physics. There are several potentially competing beyond the Standard Model mechanisms that can induce the process. It thus becomes important to disentangle the different processes. In the present study we consider the interference effect between the light left-handed and heavy right-handed Majorana neutrino exchange mechanisms. The decay rate, and consequently, the phase-space factors for the interference term are derived, based on the left-right symmetric model. The numerical values for the interference phase-space factors for several nuclides are calculated, taking into consideration the relativistic Coulomb distortion of the electron wave function and finite-size of the nucleus. The variation of the interference effect with the $Q$-value of the process is studied.       
\end{abstract}
\maketitle
\section{Introduction}
Neutrinoless double beta decay ($\NDBD$) is a process where a  nuclide decays to its isobar having two more protons. In the process two electrons are emitted without any anti-neutrinos,
\begin{align}\label{decays}
\ce{^{A}_{Z}X}\rightarrow\ce  {^{A}_{Z+2}X}+2e^-. 
\end{align}
Unlike the two neutrino double beta decay ($\TNDBD$) where two anti-neutrinos are also emitted, in $\NDBD$ the lepton number is violated by two units. In the Standard Model (SM) the lepton number is conserved. Thus the experimental observation of $\NDBD$ would indicate physics beyond the Standard Model (BSM). Moreover, the $\NDBD$ process requires the neutrinos to be Majorana fermions \cite{PhysRevD.25.2951}. All these features make $\NDBD$ an exciting process for testing BSM physics, and several BSM mechanisms are proposed for contributing to the decay \cite{PhysRevC.87.014320, 0034-4885-75-10-106301}. 

The most studied mechanism in the $\NDBD$ decay literature is the so called `standard mass mechanism' of light left-handed (LH) neutrino  exchange \cite{PhysRevC.87.014320}. However, if right-handed (RH) $V+A$ currents are also considered alongside the left-handed $V-A$ currents, then several competing mechanisms could contribute significantly to the $\NDBD$ decay rate \cite{Doi+Kotani1985}. The existence of $V+A$ currents is well motivated by new physics of BSM. One such popular BSM scenario is the left-right symmetric model (LRSM) \cite{PhysRevLett.44.912, Pati:1974yy}, which is presently actively investigated at LHC \cite{Khachatryan:2014dka}. RH currents are naturally considered in this model, since parity is restored at high energies. The extended gauge group for the electroweak sector of the model is $SU(2)_L\times SU(2)_R\times U(1)_Y$, where RH neutrinos are incorporated as part of the $SU(2)_R$ doublet. LRSM provides a natural framework for type I \cite{PhysRevLett.44.912} and type II \cite{Mohapatra+Senjanovic1981} seesaw mechanisms, which explains the smallness of neutrino mass. Apart from being crucial in determining absolute values of neutrino masses, $\NDBD$ is an important low energy probe to study high energy BSM mechanisms \cite{Bilenky+Giunti2015}. Invariably this also requires disentangling the underlying mechanism inducing the process \cite{PhysRevD.83.113003}. 

Because of the existence of RH currents there will be several competing contributions to the decay rate of $\NDBD$ in the LRSM scenario \cite{Barry:2013xxa}. Additionally, the seesaw mechanism requires the existence of heavy, mostly sterile neutrinos \cite{Mohapatra+Senjanovic1981}. Neutrino mixing schemes would then naturally incorporate heavy mass eigenstates for both LH and RH neutrinos (see below for details). The inverse half-life formula for $\NDBD$ has the following general structure,
\begin{align}\label{genhalf-life}
\[T_{\sfrac{1}{2}}^{0\nu}\]^{-1}=\abs{\sum_{i}\Big(\text{NPP $\&$ PPP}\Big)_i\times\Big(\text{PSF}\Big)_i^{\sfrac{1}{2}}\times\Big(\text{NME}\Big)_i}^2,
\end{align}
where NPP and PPP are the neutrino and particle physics parameters arising from the BSM physics. The phase-space factors (PSF) take into account the kinematical factors of the two outgoing electrons. 
The summation $i$ is over all possible mechanisms that could induce the $\NDBD$ process. Because of the modulus squared, interference between different mechanisms also contribute to the total decay rate of the process. 
For a long time the PSF were generally considered as being reliably calculated \cite{Doi+Kotani1985,SuhonenCivitarese1998}, but recent re-evaluations \cite{PhysRevC.85.034316,StoicaMirea2013,Mirea2015,Stefanik+Dvornicky2015,Neacsu:2015uja} have shown diferences of up to 100\% in the case of heavier nuclei (e.g. $G^{0\nu}_{08}$ of $^{150}$Nd), when comparing to the older results. Presently, the uncertainties in the PSF are maintained under control and new results are in agreement to each other.
Lastly, the NME are the relevant nuclear matrix elements for the nuclear transition between the initial and final nuclei. There are several methods comonly used to calculate the NME, of which the most employed ones are the interacting shell model (ISM) \cite{Retamosa1995,HoroiStoicaBrown2007,HoroiStoica2010,PhysRevC.87.014320,Horoi:2015gdv,HoroiBrown2013,Caurier2008,MenendezPovesCaurier2009}, the quasi-random phase approximation (QRPA) \cite{SuhonenCivitarese1998,Suhonen2011,Suhonen2015,Simkovic2008,Simkovic2009,SimkovicRodin2013}, the interacting-boson approach (IBM2) \cite{Barea2013,Barea2015}, the projected Hartree-Fock-Bogoliubov (PHFB) \cite{Rath2013}, and the generator coordinator method (GCM) \cite{Rodriguez2010,Song2014}. The current situation of the NME is not yet settled, with differences of up to 300\% among the results of different groups and methods (See Fig. 6 and Fig. 7 of \cite{NeacsuHoroi2016}). Also, relatively few NME results other than those for the 'standard mass mechanism' exist.

If two competing mechanisms are considered for the $\NDBD$ decay, as is the case for our present study, then the magnitude of the contribution of the interference to the half-life will primarily depends on the PSF. The phase difference arising from the two different complex neutrino physics parameters will act as the phase of the interference oscillation. The chiral structure of the outgoing electrons  determines the nature of the PSF for the individual mechanisms. This  also results in a different form of the PSF for the interference. The total decay rate for the process is obtained by integrating the PSF over the energies of the outgoing electrons. Several effects must be considered in order to accurately determine numerical values for the PSF. The Coulomb attraction of the daughter nucleus will distort the wave functions of the outgoing electrons \cite{BUHRING1963472}. Moreover, the finite size of the nucleus will affect the distorted electron wavefunction \cite{PhysRevC.85.034316, Stefanik+Dvornicky2015}.

The subject of interference between pairs of $\NDBD$ mechanisms is obscure in recent literature, with the latest paper devoted to this subject being 34 years old \cite{HALPRIN1983335}. Since there is great interest to disentangle the different underlying mechanisms of $\NDBD$, it is important that the interference between mechanisms is also considered. In the present paper we re-consider the interference between the light and heavy Majorana neutrinos. We calculate the decay rate of $\NDBD$ for light and heavy neutrino exchange in LRSM. 
%The light neutrinos are predominantly assumed from the LH fields. The heavy neutrinos are predominantly from RH fields. 
The focus of our paper is to investigate the contribution of interference effects. We revisit the calculations and include relativistic Fermi function and finite-size effect of the nucleus in the phase-space factors (PSF) of the interference terms. We numerically calculate the interference PSF for different nuclides and compare our results with those of Ref. \cite{HALPRIN1983335}. Although the numerical results of \cite{HALPRIN1983335} are similar to ours, the formulae used in Ref. \cite{HALPRIN1983335} seems to contain typos that make them difficult to use. In addition, our analysis is extended to more isotopes than previously considered and for transitions to excited states. The paper is organized as follows: Section II presents the general formalism for $\NDBD$ considering both LH and RH currents. We derive the exact expressions for the PSFs. Section III presents the numerical results for the PSFs and the effect of interference for different nuclides considering transitions to the ground state (g.s.) and to the first excited $0^+$ state of the daughter nucleus.

\section{Formalism for $\NDBD$ decay}
The derivation of the decay rate for $\NDBD$ begins with the effective current-current Hamiltonian incorporating the RH leptonic and hadronic currents \cite{Doi+Kotani1985},
\begin{align}\label{Hbeta}
\Hbeta=\ferm 2\[&\(\elL\ga\nu'_{eL}\)\(J^{\dagger}_{L\alpha}+\kappa J^{\dagger}_{R\alpha}\)\nn\\
&+\(\elR\ga\nu'_{eR}\)\(\eta J^{\dagger}_{L\alpha}+\lambda J^{\dagger}_{R\alpha}\)+\text{H.c.}\].
\end{align}
Here $e_{L(R)}$ and $\nu'_{eL(R)}$ are the LH(RH) electron and electron-neutrino fields, respectively. $J_{L(R)\alpha}$ is the charged LH (RH) hadronic current. $\lambda$, $\eta$ and $\kappa$ are the respective coupling constants between $(V+A)(V+A)$, $(V+A)(V-A)$ and $(V-A)(V+A)$ interactions. $G_F$ and $\theta_C$ are the Fermi constant and Cabbibo angle, respectively. H.c. denotes the Hermitian conjugate. 

\begin{figure}[H]
\begin{minipage}{0.5\textwidth}
\centering
\subcaptionbox{Light neutrino exchange.\label{Left-light}}
{\includegraphics[scale=0.8]{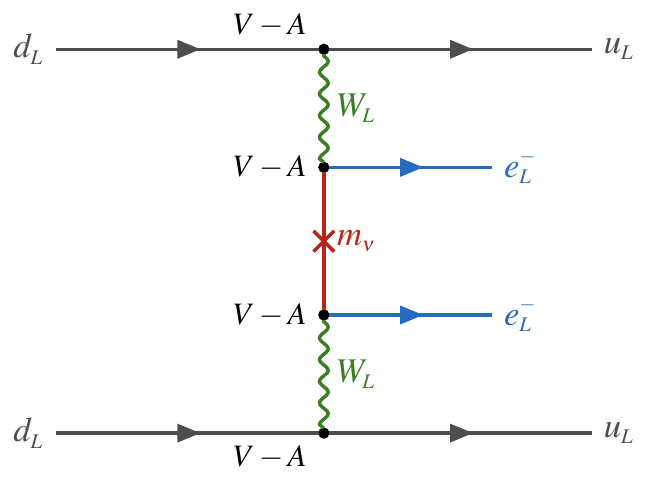}}
\end{minipage}
\begin{minipage}{0.5\textwidth}
\centering
\subcaptionbox{Heavy neutrino exchange.\label{Right-heavy}}
{\includegraphics[scale=0.8]{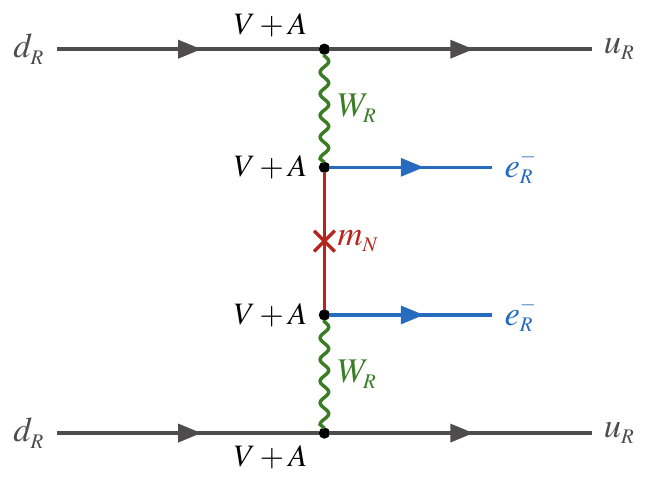}}
\end{minipage}
\caption{Light and heavy neutrino exchange mechanisms for $\NDBD$ in LRSM.}\label{mechanisms}
\end{figure}

In the left-right symmetric model (LRSM) the electron-neutrino mixing is given as \cite{Horoi:2015gdv}:
\begin{align}
 &\nu'_{eL}=\sum_k^{\text{light}}U_{ek}\nu_{kL}+\sum_k^{\text{heavy}}S_{ek}N^c_{kR},\label{Lmix}\\
 &\nu'_{eR}=\sum_k^{\text{light}}T^*_{ek}\nu^c_{kL}+\sum_k^{\text{heavy}}V^*_{ek}N_{kR}.\label{Rmix}
\end{align}
Here the flavor eigenstates $\nu'$ are linear combinations of light ($\nu$) and heavy ($N$) mass eigenstates. The mixing matrices $U$ and $V$ are almost unitary while $S$ and $T$ are taken to be small. Thus the LH fields are predominantly light neutrinos, whereas the RH fields are predominantly heavy, $\nu'_{eL}\approx \sum_k^{\text{light}}U_{ek}\nu_{kL}$ and $\nu'_{eR}\approx\sum_k^{\text{heavy}}V^*_{ek}N_{kR}$. Thus under these assumptions, the two dominating diagrams for the second order process of $\NDBD$ arising from the Hamiltonian of \eq{Hbeta} are shown in Figure \ref{mechanisms}. Figure \ref{Left-light} is the standard mass mechanism of light neutrino, $m_{\nu}$, exchange. In Figure \ref{Right-heavy}, $\NDBD$ is induced by the exchange of heavy neutrino, $m_N$. The coupling constant between $(V+A)(V+A)$ interaction, $\lambda$, is shown to be  $(m_{W_L}/m_{W_R})^2$ in the LRSM. Here $m_{W_L/R}$ is the mass of the LH (RH) W boson. 

Applying second order perturbation theory to our interaction Hamiltonian, \eq{Hbeta}, we get the S-matrix for $\NDBD$ process as follows,
\begin{align}\label{fullmel}
&\matrixel{f}{S^{(2)}}{i}^{0\nu}=iA_0\left[\eta^{\nu}_L\,\EL\,M^{\nu}+\eta_R^{N}\,\ER\,M^{N}\right]2\pi\delta(E_f+E_1+E_2-E_i).
\end{align}
The first and the second terms inside the square brackets correspond to the light and heavy neutrinos, respectively.  The dimensionful constant $A_0$ is defined as:
\begin{align}
&A_{0}=\gAA\fermi\frac{1}{r_A},\label{A0}
\end{align}
where $g_A=1.27$ is the axial vector constant \cite{Towner+Hardy1995}, $m_e$ is the electronic mass, and $r_A=m_er_0A^{\frac{1}{3}}$ with $r_0=1.2$ fm, is the nuclear radius in the units of $m_e$, $A$ being the mass number of the nuclide. The neutrino physics parameters for the light and heavy neutrinos are defined respectively as follows:    
\begin{align}
&\eta_L^{\nu}=\light,\\
&\eta_R^{N}=\heavy.
\end{align}
Here $m_k$ and $M_k$ are the masses of the light and heavy neutrinos, respectively. Similarly, the Majorana phase factors for light and heavy neutrinos are $\xi_k$ and $\Xi_k$, respectively. The proton mass $m_p$ is introduced for dimensional consistency. The nuclear matrix elements (NME) for the light ($M^{\nu}$) and heavy ($M^N$) neutrinos are the combined Fermi, Gamow-Teller and tensor matrix elements as defined in \cite{Simkovic+Pantis1999}: 
\begin{align}
&M^{\nu/N}=\[M_{GT}^{\nu/N}-\left(\frac{g_V}{g_A}\right)^2M_F^{\nu/N}+M_T^{\nu/N}\].
\end{align}
The delta function imposes the energy conservation, $E_i$ and $E_f$ are the energies of the initial and final nuclei, respectively. $E_1$ and $E_2$ are the energies of the two outgoing electrons. $\EL$ and $\ER$ correspond to the two final state electron current for light and heavy neutrino exchange, respectively: 
\begin{align}
&\EL=\[\bar u(p_1s_1)(1+\gamma^5)C\bar u^T(p_2s_2)\],\\
&\ER=\[\bar u(p_1s_1)(1-\gamma^5)C\bar u^T(p_2s_2)\],
\end{align}
where the free particle spinor $u$ is defined for the two outgoing electrons having momenta $p_1$, $p_2$ and spins $s_1$, $s_2$, respectively. The presence of the $\gf$ matrices makes the chiral structure of the two currents manifestly different, namely both the electrons are LH for the case of light neutrino exchange, whereas both neutrinos are RH for the heavy neutrino exchange (see Figure \ref{mechanisms}). The amplitude of the process is:
\begin{align}
\mathcal{M}_{fi}^{0\nu}=A_0\eta^{\nu}_L\,\EL\,M^{\nu}+A_0\eta_R^{N}\,\ER\,M^{N}=\mathcal{M}_{fi}^{\text{\tiny{light}}}+\mathcal{M}_{fi}^{\text{\tiny{heavy}}}.
\end{align}
Squaring the amplitude one gets the interference terms between the light and heavy neutrino exchange mechanisms,  
\begin{align}
\abs{\mathcal{M}_{fi}^{0\nu}}^2&=\abs{\mathcal{M}_{fi}^{\text{\tiny{light}}}}^2+\abs{\mathcal{M}_{fi}^{\text{\tiny{heavy}}}}^2
+\mathcal{M}_{fi}^{\text{\tiny{light}}}\mathcal{M}_{fi}^{\text{\tiny{heavy}}*}+\mathcal{M}_{fi}^{\text{\tiny{light}}*}\mathcal{M}_{fi}^{\text{\tiny{heavy}}}.
\end{align}
The first and second terms are the amplitude squares for the light and heavy neutrinos, respectively. The third and fourth terms are the interference terms between the light and heavy neutrinos. The spin averaged amplitude square is then,
\begin{align}
\abs{\xbar{\mathcal{M}_{fi}^{0\nu}}}^2&=A_0^2\sum_{\text{spin}}\[|\eta_L^{\nu}|^2|M^{\nu}|^2|\EL|^2+|\eta_R^{N}|^2|M^{N}|^2|\ER|^2
\nn\\
&+\eta_L^{\nu}(\eta_R^{N})^*M^{\nu}(M^{N})^*\EL\ER^*+(\eta_L^{\nu})^*\eta_R^{N}(M^{\nu})^*M^{N}\EL^*\ER\].\label{spinavgamp}
\end{align}
For $\NDBD$ we only need to sum over the two outgoing electrons. Calculating the spin sums using Casimir's trick and trace theorems \cite{halzen1984quarks} we get:
\begin{align}
&\sum_{\substack{\text{$e_1$,$e_2$}\\ \text{spin}}}\left|\EL\right|^2=\!\!\!\sum_{\substack{\text{$e_1$,$e_2$}\\ \text{spin}}}\left|\ER\right|^2=8E_1E_2\left(1-\frac{\scmag{p}{1}\scmag{p}{2}}{E_1E_2}\cos\theta\right),\\
&\sum_{\substack{\text{$e_1$,$e_2$}\\ \text{spin}}}\EL\ER^*=\!\!\!\sum_{\substack{\text{$e_1$,$e_2$}\\ \text{spin}}}\EL^*\ER=-8m_e^2.
\end{align}
Here $\vc{p}_{1}$ and $\vc{p}_{2}$ are the three-momenta of the two electrons emitted at an angle $\theta$. Thus we see that the spin average for the interference term in the amplitude square is different from that of the terms for the individual mechanisms. For the individual terms, the spin average gives a contribution, which depends on the energies of the electrons while for the interference term, the spin average is independent of the electronic energies. The phase-space factors for the individual and interference terms will be different due to this difference in the spin average.    

The total decay rate of the process is then obtained using the Fermi's Golden rule \cite{halzen1984quarks} ($\hbar=1$, $c=1$), and integrating over the phase-space of the two electrons,
\begin{align}\label{Gamma}
\Gamma^{0\nu}=\frac{1}{2!}2\pi\int\denstat\abs{\xbar{\mathcal{M}_{fi}^{0\nu}}}^2\delta(E_f+E_1+E_2-E_i). 
\end{align}
Carrying out the angular integration for the 3D momentum integral in polar coordinates we can then re-express $\scmag{p}{k}$ in terms of relativistic energy $E_k=\vc{p}_k^2+m_e^2$, where $k=1,2$. The $E_2$ integral then can be carried out through the $\delta$-function. Due to energy conservation, we can express $E_2$ in terms of $E_1$. Since the kinetic energy is shared between the two electrons, the $E_1$ integral is carried out between the limits $m_e$ and $Q+m_e$.
%(in natural units, $\hbar=1$, $c=1$ ). 
The two outgoing electrons are attracted by the Coulomb field of the daughter nucleus. The free electron wavefunction is thus modified by the Fermi factor $F_0$ as given in reference \cite{behrens1982electron}. The total decay rate for $\NDBD$ becomes:  
\begin{align}
&\Gamma^{0\nu}=\frac{g_A^4(G_F\,\cos\,\theta_c)^4m_e^9}{(2\pi)^5r_A^2}\!\!\!\int^{T+1}_{1}\!\!\!\dd{\epsilon_1}\tilde{p}_1\tilde{p}_2F_0(Z_s,\epsilon_1)F_0(Z_s,\epsilon_2)\nn\\
&\times\Bigg[\epsilon_1\epsilon_2 \[|\eta_L^{\nu}|^2|M^{\nu}|^2+|\eta_R^{N}|^2|M^{N}|^2\]-\[\eta_L^{\nu}\eta_R^{N*}M^{\nu}M^{N*}+\eta_L^{\nu*}\eta_R^{N}M^{\nu*}M^{N}\]\Bigg],
\end{align}
where we have switched to energies ($\epsilon_k$) and momenta ($\tilde{p}_k$) in the $m_e$ units. Similarly, $T$ is the $Q$-value in the $m_e$ units. The half life is then related to the total decay rate by:
\begin{align}\label{halflife}
\[T_{\sfrac{1}{2}}^{0\nu}\]^{-1}&=g_A^4G_{1}^{0\nu}\[|\eta_L^{\nu}|^2|M^{\nu}|^2+|\eta_R^{N}|^2|M^{N}|^2\]\nn\\
&-g_A^4G_{1}^{0\nu '}\[\eta_L^{\nu}\eta_R^{N*}M^{\nu}M^{N*}+\eta_L^{\nu*}\eta_R^{N}M^{\nu*}M^{N}\], 
\end{align}
where the integrated phase-space factors (PSF) are defined by:
\begin{align}
&G_{1}^{0\nu}=\frac{g^{0\nu}}{r_A^2}\!\!\!\int^{T+1}_{1}\dd{\epsilon_1}F_0(Z_s,\epsilon_1)F_0(Z_s,\epsilon_2)\tilde{p}_1\tilde{p}_2\epsilon_1\epsilon_2\delta(\epsilon_2+\epsilon_1-T-2),\label{G1}\\
&G_{1}^{0\nu '}=\frac{g^{0\nu}}{r_A^2}\!\!\!\int^{T+1}_{1}\dd{\epsilon_1}F_0(Z_s,\epsilon_1)F_0(Z_s,\epsilon_2)\tilde{p}_1\tilde{p}_2\delta(\epsilon_2+\epsilon_1-T-2),\label{G1'}
\end{align}
with the common dimensionful constant having the value,
\begin{align}
g^{0\nu}=\gOnu=2.8\times10^{-22}\,\,\, \text{yr}^{-1}.
\end{align}
The PSF for the light and heavy neutrino exchange are the same, $G_{1}^{0\nu}$ of \eq{G1}. As \eq{G1} and (\rf{G1'}) show, the PSF for the interference term, $G_{1}^{0\nu '}$, has a different structure than the PSF of the individual mechanisms,  $G_{1}^{0\nu}$. Because of the absence of electron energies, the interference PSF is suppressed considerably compared to the `non-interference' PSF.

\section{Results}
The accuracy of PSF calculations depends on certain assumptions and methods. References \cite{PhysRevC.85.034316} and \cite{Stefanik+Dvornicky2015} considered the effect of finite nuclear size and screening of nuclear charge due to atomic electrons in calculating the non-interference PSF. An easy to use, faster and sufficiently accurate method was recently introduced in \cite{Neacsu:2015uja} by considering a screening factor $S_f$ to the charge of the final nucleus $Z_f$, still retaining the original assumption of point-like nuclear charge. This modification to the charge ($Z_s=\frac{S_f}{100}Z_f$) replicates the effects of finite nuclear size and electron screening to good accuracy. For our case we consider the value $S_f=94.5\%$ for $G_1^{0\nu}$ (see table IV of \cite{Neacsu:2015uja}).
\begin{table}[H]
\centering
\caption{Numerical values of PSF of different nuclei for g.s.$\rightarrow$g.s. transitions. Column 4 lists the values of the interference PSF \eq{G1'} and column 5 is the ratio to the individual PSF in $\%$ (see \eq{G1}). ($Q_{\beta\beta}$ values are taken from \cite{Stefanik+Dvornicky2015}).\label{PSF-table}}
\begin{tabular}{ccccc}
\toprule
\multicolumn{1}{c}{Nuclei}    & \vtop{\hbox{\strut \,\,\,$Q_{\beta\beta}$}\hbox{\strut[MeV]}}   & \vtop{\hbox{\strut\qquad $G^{0\nu}_1$}\hbox{\strut [$\times10^{-14}$ $\text{yr}^{-1}$]}}  &  \vtop{\hbox{\strut\qquad $G^{0\nu '}_1$}\hbox{\strut [$\times10^{-15}$ $\text{yr}^{-1}$]}} & \vtop{\hbox{\strut $\varepsilon=\frac{G_{1}^{0\nu'}}{G_{1}^{0\nu}}$}\hbox{\strut \,\,\,\,[$\%$]}}  \\ \midrule
\ce{^{48}Ca}  			& $4.27226$  & $2.45462$                                    								     & $1.09027$                              									 & $4.44168$                                            							  \\
\ce{^{76}Ge}  			& $2.03904$  & $0.228003$                                    								     & $0.28987$                               									 & $12.7134$                                                                                                \\
\ce{^{82}Se}  			 & $2.99512$ & $0.996509$                                    								     & $0.757323$                              									 & $7.59976$                                            							  \\
\ce{^{96}Zr}  			 & $3.35037$ & $2.04454$                                    								     & $1.32482$                              									 & $6.47976$                                                                                                \\
\ce{^{100}Mo} 			  & $3.03440$ & $1.57402$                                    								     & $1.17824$                              									 & $7.48553$                                            							  \\
\ce{^{110}Pd} 			  & $2.01785$ & $0.465953$                                    								     & $0.603145$                              									 & $12.9443$                                            							  \\
\ce{^{116}Cd} 			  & $2.8135$ & $1.65694$                                   								     & $1.38323$                              									 & $8.34808$                                            							  \\
\ce{^{124}Sn} 			 & $2.28697$ & $0.886628$                                    								     & $0.979603$                              									 & $11.0486$                                            							  \\
\ce{^{128}Te} 			  & $0.86795$ & $0.0554$                                     								     & $0.017355$                              									 & $31.3251$                                            							  \\
\ce{^{130}Te} 			  & $2.5269$ & $1.4104$                                     								     & $1.36624$                              									 & $9.68692$                                            							  \\
\ce{^{136}Xe} 			  & $2.45783$ & $1.44863$                                    								     & $1.45738$                              									 & $10.0604$                                            							  \\
\ce{^{150}Nd} 			  & $3.37138$ & $6.60043$                                    								     & $4.27367$                              									 & $6.47483$                                            							  \\ \bottomrule
\end{tabular}
\end{table}
\begin{table}[H]
\centering
\caption{Same as Table 1, for transitions from g.s.$\rightarrow$ $1^{st}$ $0^{+}$ excited states. ($Q_{\beta\beta}$ values are taken from \cite{Neacsu:2015uja}). \label{PSF1-table}}
\begin{tabular}{ccccc}
\toprule
\multicolumn{1}{c}{Nuclei}    & \vtop{\hbox{\strut \,\,\,$Q_{\beta\beta}$}\hbox{\strut[MeV]}}   & \vtop{\hbox{\strut\qquad $G^{0\nu}_1$}\hbox{\strut [$\times10^{-15}$ $\text{yr}^{-1}$]}}  &  \vtop{\hbox{\strut\qquad $G^{0\nu '}_1$}\hbox{\strut [$\times10^{-16}$ $\text{yr}^{-1}$]}} & \vtop{\hbox{\strut $\varepsilon=\frac{G_{1}^{0\nu'}}{G_{1}^{0\nu}}$}\hbox{\strut \,\,\,\,[$\%$]}}  \\ \midrule
\ce{^{48}Ca}  			& $1.275$  & $0.292636$                                    								     & $0.633955$                              									 & $21.6636$                                            							  \\
\ce{^{76}Ge}  			& $0.917$  & $0.185314$                                    								     & $0.552101$                               									 & $29.7927$                                                                                                \\
\ce{^{82}Se}  			 & $1.508$ & $0.906993$                                    								     & $1.64925$                              									 & $18.1837$                                            							  \\
\ce{^{96}Zr}  			 & $2.202$ & $4.43543$                                    								     & $5.12937$                              									 & $11.5646$                                                                                                \\
\ce{^{100}Mo} 			  & $1.904$ & $3.03925$                                    								     & $4.221$                              									 & $13.8883$                                            							  \\
\ce{^{110}Pd} 			  & $0.5472$ & $0.120312$                                    								     & $0.531163$                              									 & $44.1487$                                            							  \\
\ce{^{116}Cd} 			  & $1.057$ & $0.720768$                                   								     & $1.89277$                              									 & $26.2605$                                            							  \\
\ce{^{124}Sn} 			 & $1.12$ & $0.953037$                                    								     & $2.36911$                              									 & $24.8585$                                            							  \\
\ce{^{130}Te} 			  & $0.7335$ & $0.358299$                                     								     & $1.28598$                              									 & $35.8911$                                            							  \\
\ce{^{136}Xe} 			  & $0.879$ & $0.651285$                                    								     & $2.0187$                              									 & $30.9956$                                            							  \\
\ce{^{150}Nd} 			  & $2.631$ & $27.5308$                                    								     & $25.3183$                              									 & $9.19636$                                            							  \\ \bottomrule
\end{tabular}
\end{table}

\begin{figure}[H]
\centering
\includegraphics[scale=0.5]{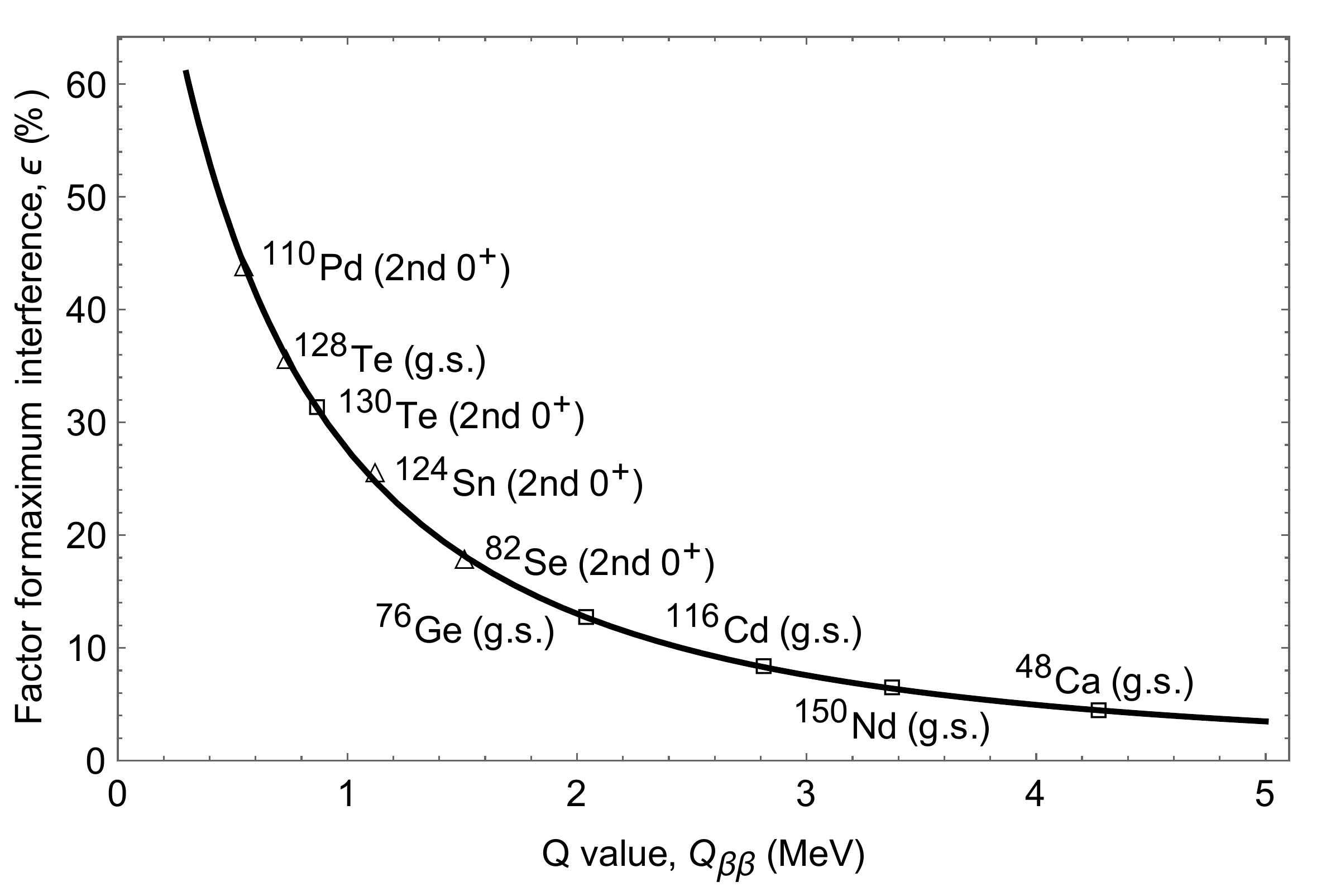}
\caption{Variation of the factor for maximum interference $\varepsilon$ with $Q_{\beta\beta}$. Plot is obtained by varying the $Q$-value, while keeping the mass fixed to $\ce{^{76}Ge}$ and the charge of the final nucleus $Z_f=34$ of $\ce{^{76}Se}$. Different nuclei are added in the plot for the g.s. and first $0^+$ excited states $Q_{\beta\beta}$ values.\label{plot}}
\end{figure}

The effect of the  interference term was considered in \cite{HALPRIN1983335} where the numerical values of the suppression factors of the term for different nuclei were calculated. We try to verify the values claimed in \cite{HALPRIN1983335} based on our derivation of the interference term PSF (see \eq{G1'}). We analyze the contribution of the PSF for maximum interference so as to make our conclusions as general as possible. This is in anticipation of its smallness, as was already claimed in \cite{HALPRIN1983335} and \cite{PhysRevD.83.113003}. To make the analysis for maximum interference more transparent we introduce certain assumptions and notations. The NME are taken to be real $M^{\nu*}=M^{\nu}$, $M^{N*}=M^N$. The LNV parameters are nonetheless treated as complex having the phases, 
\begin{align}
\eta_L^{\nu}=\abs{\eta_L^{\nu}}\e^{i\phi_1}\qquad,\qquad\eta_R^N=\abs{\eta_R^N}\e^{i\phi_2}.
\end{align}
Furthermore we make the assumptions $\abs{\eta_L^{\nu}}M^{\nu}\approx\abs{\eta_R^{N}}M^{N}\approx\eta M$. Thus the inverse half-life of \eq{halflife} can be rewritten so as to give maximum interference,
\begin{align}
&\[T_{\sfrac{1}{2}}^{0\nu}\]^{-1}\!\!\!\!\!\!\!\!\!\approx g_A^4\left[G_{1}^{0\nu}\!\!\left[\abs{\eta}^2\abs{M}^2+\abs{\eta}^2\abs{M}^2\right]-G_{1}^{0\nu'}\!\!\left[\abs{\eta}^2\abs{M}^2\e^{(\phi_1-\phi_2)}+\abs{\eta}^2\abs{M}^2\e^{(\phi_2-\phi_1)}\right]\right]\nn\\ 
             &=2g_A^4G_{1}^{0\nu}\abs{\eta}^2\abs{M}^2\left[1-G_{1}^{0\nu'}/G_{1}^{0\nu}\cos{\phi}\right]=2g_A^4G_{1}^{0\nu}\abs{\eta}^2\abs{M}^2\left[1-\varepsilon\cos{\phi}\right],\label{interhalflife}
\end{align}
where we have defined the phase difference between the two LNV parameters as $\phi=\phi_1-\phi_2$ and the ratio between the two phase space factors of \eq{G1}, (\rf{G1'}) as $\varepsilon=G_{1}^{0\nu'}/G_{1}^{0\nu}$. Depending on the value of $\varepsilon$, the $\varepsilon\cos{\phi}$ term will mostly determine the contribution of the interference. The numerical values are tabulated in Tables \ref{PSF-table} and \ref{PSF1-table} for the g.s. $\rightarrow$ g.s. and g.s. $\rightarrow$ $0^+$ ($1^{st}$ excited) transitions, respectively. The first excited $0^+$ transitions are also considered because of their potential experimental relevance. 
%The first excited $0^+$  transitions would emit gamma rays for subsequent de-excitations to the ground states. These gamma rays could then be observed during the $\NDBD$ decay to the first excited $0^+$ state of the final nuclide. 
The fifth column in both tables lists the values of the factor for maximum interference, $\varepsilon$ in $\%$. From the values we can  clearly see that the interference effect decreases with the $Q_{\beta\beta}$ values. From Table \ref{PSF-table} one observes quite clearly the extreme cases for $\ce{^{48}Ca}$ and $\ce{^{128}Te}$. A 4$\%$ increase to the $\NDBD$ decay rate due to consideration of the interference effect for a $Q_{\beta\beta}$ value of 4.27 MeV for $\ce{^{48}Ca}$, whereas for $\ce{^{128}Te}$ the contribution is as high as 31$\%$ for a $Q_{\beta\beta}$ of 0.868 MeV. The variation of $\varepsilon$ with $Q_{\beta\beta}$ is plotted in Figure \ref{plot}, obtained by fixing the mass number $A=76$ for Ge and the proton number of the final nucleus $Z_f=34$. As the plot suggests, for decay modes of smaller $Q_{\beta\beta}$ values the effect of interference will be stronger. The effect can be as high as $\approx$ 50$\%$ as can be seen for the first excited $0^+$ transition for $\ce{^{110}Pd}$ (see Table \ref{PSF1-table} and Figure \ref{plot}). We also studied the variation of $\varepsilon$ with mass number $A$ and proton number $Z_f$ for the final nucleus. No significant dependence on $A$ and $Z_f$ was found.     
The exchange of heavy neutrinos for inducing $\NDBD$ was first considered in \cite{PhysRevD.13.2567}. The interference between light and heavy neutrinos was then considered in \cite{HALPRIN1983335}. It was found that the PSF for the interference term was typically factor of 10 smaller than the PSF of the individual mechanisms \cite{HALPRIN1983335, PhysRevD.83.113003}. The suppression factors of the interference term were calculated for several nuclides and was found to be less than 10$\%$. The smallness of the contribution of the interference is mainly due to the opposite chiral structures of the outgoing electrons. Due to the Coulomb attraction of the daughter nucleus, the outgoing electron wave functions are distorted. Therefore the PSF has to be calculated (numerically) taking this into account by introducing Fermi factor \cite{BUHRING1963472}. In \cite{HALPRIN1983335} this was done by using non-relativistic Fermi factor, which is independent of the $Q$-value of the process. Moreover the electrons were assumed to be ultra-relativistic in the analysis in order to arrive at the numerical values of the small suppression factors of the interference term. This is in contrast to the consideration of the non-relativistic Fermi factor for electron wave function used in Ref. \cite{HALPRIN1983335}. For our analysis we have correctly considered the relativistic Fermi factor. In addition we have also considered the effect of finite nuclear size. Although the numerical results obtained in our current study are very close to the values in references \cite{HALPRIN1983335} and \cite{PhysRevD.83.113003}, our results are more general since the assumption of ultra-relativistic electrons can be relaxed. Consideration for relativistic Fermi factor and finite nuclear size extend the analysis and allows us to predict the $Q_{\beta\beta}$ values for which the effect of interference can be observable.

\section{Conclusions}

In summary, we studied the interference effects to the $0\nu\beta\beta$ decay rate when contributions from the light left-handed and heavy right-handed neutrino exchange mechanisms are considered. These effects were first analyzed long time ago in Ref. \cite{HALPRIN1983335} under some simplifying assumptions, a simple relation for the relative interference amplitude was presented and numerical values for few isotopes were provided (see also \cite{PhysRevD.83.113003}). The general conclusion was that these effects are small and can be neglected.  Unfortunately, the analytical expression seem to be marred by typos and one needed to redo the analysis to extend it to other isotopes of recent experimental interest. 
In addition, for a long time the standard mass mechanism was the only one mainly considered, and the results of Ref. \cite{HALPRIN1983335} were almost forgotten. 

In recent years, however, the contributions from other mechanisms, especially those related to the LRSM, became relevant and competitive to BSM studies at LHC and elsewhere. In this letter we extended the analysis of Ref. \cite{HALPRIN1983335} by considering the relativistic distortion of the outgoing electrons wave functions, the finite size effects of the daughter nucleus, and by applying the new formalism to all isotopes of recent experimental interest. In addition, we provide an analysis of the relative interference factor as a function of $Q_{\beta\beta}$, mass number A, and charge of the daughter $Z_f$, and we find that only its decrease with $Q_{\beta\beta}$ is relevant. This feature indicates that the relative interference factor might not be negligible for cases where $Q_{\beta\beta}$ is small, such as that of $^{128}$Te and for the transitions to the first excited $0^+$ states (e.g. it reaches 44\% for $^{110}$Pa). Therefore, we provide numerical results for all these new transitions that could be of experimental interest.

Finally, the analysis presented can be extended to other pairs of $\NDBD$ mechanisms where both outgoing electrons have different helicities. Examples of such mechanisms described in within the effective field theory approach can be found in Ref. \cite{Deppisch2012}. 

\section*{Acknowledgments}

Support from the US NSF Grant no. PHY-1404442 and the NUCLEI SciDAC Collaboration under US Department of Energy Grant no. DE-SC0008529  is acknowledged. Mihai Horoi also acknowledges US Department of Energy Grant no. DE-SC0015376. \\

\bibliographystyle{elsarticle-num}

\bibliography{References_Paper}

\begin{thebibliography}{10}
\expandafter\ifx\csname url\endcsname\relax
  \def\url#1{\texttt{#1}}\fi
\expandafter\ifx\csname urlprefix\endcsname\relax\def\urlprefix{URL }\fi
\expandafter\ifx\csname href\endcsname\relax
  \def\href#1#2{#2} \def\path#1{#1}\fi

\bibitem{PhysRevD.25.2951}
J.~Schechter, J.~W.~F. Valle,
  \href{http://link.aps.org/doi/10.1103/PhysRevD.25.2951}{Neutrinoless
  double-$\ensuremath{\beta}$ decay in
  su(2)\ifmmode\times\else\texttimes\fi{}u(1) theories}, Phys. Rev. D 25 (1982)
  2951--2954.
\newblock \href {http://dx.doi.org/10.1103/PhysRevD.25.2951}
  {\path{doi:10.1103/PhysRevD.25.2951}}.
\newline\urlprefix\url{http://link.aps.org/doi/10.1103/PhysRevD.25.2951}

\bibitem{PhysRevC.87.014320}
M.~Horoi, \href{http://link.aps.org/doi/10.1103/PhysRevC.87.014320}{Shell model
  analysis of competing contributions to the double-$\ensuremath{\beta}$ decay
  of ${}^{48}$ca}, Phys. Rev. C 87 (2013) 014320.
\newblock \href {http://dx.doi.org/10.1103/PhysRevC.87.014320}
  {\path{doi:10.1103/PhysRevC.87.014320}}.
\newline\urlprefix\url{http://link.aps.org/doi/10.1103/PhysRevC.87.014320}

\bibitem{0034-4885-75-10-106301}
J.~D. Vergados, H.~Ejiri, F.~Šimkovic,
  \href{http://stacks.iop.org/0034-4885/75/i=10/a=106301}{Theory of
  neutrinoless double-beta decay}, Reports on Progress in Physics 75~(10)
  (2012) 106301.
\newline\urlprefix\url{http://stacks.iop.org/0034-4885/75/i=10/a=106301}

\bibitem{Doi+Kotani1985}
M.~Doi, T.~Kotani, E.~Takasugi,
  \href{http://ptps.oxfordjournals.org/content/83/1.abstract}{Double beta decay
  and majorana neutrino}, Progress of Theoretical Physics Supplement 83 (1985)
  1--175.
\newblock \href {http://dx.doi.org/10.1143/PTPS.83.1}
  {\path{doi:10.1143/PTPS.83.1}}.
\newline\urlprefix\url{http://ptps.oxfordjournals.org/content/83/1.abstract}

\bibitem{PhysRevLett.44.912}
R.~N. Mohapatra, G.~Senjanovi\ifmmode~\acute{c}\else \'{c}\fi{},
  \href{http://link.aps.org/doi/10.1103/PhysRevLett.44.912}{Neutrino mass and
  spontaneous parity nonconservation}, Phys. Rev. Lett. 44 (1980) 912--915.
\newblock \href {http://dx.doi.org/10.1103/PhysRevLett.44.912}
  {\path{doi:10.1103/PhysRevLett.44.912}}.
\newline\urlprefix\url{http://link.aps.org/doi/10.1103/PhysRevLett.44.912}

\bibitem{Pati:1974yy}
J.~C. Pati, A.~Salam, {Lepton Number as the Fourth Color}, Phys. Rev. D10
  (1974) 275--289, [Erratum: Phys. Rev.D11,703(1975)].
\newblock \href {http://dx.doi.org/10.1103/PhysRevD.10.275,
  10.1103/PhysRevD.11.703.2} {\path{doi:10.1103/PhysRevD.10.275,
  10.1103/PhysRevD.11.703.2}}.

\bibitem{Khachatryan:2014dka}
V.~Khachatryan, et~al., {Search for heavy neutrinos and $\mathrm {W}$ bosons
  with right-handed couplings in proton-proton collisions at $\sqrt{s} =
  8\,\text {TeV} $}, Eur. Phys. J. C74~(11) (2014) 3149.
\newblock \href {http://arxiv.org/abs/1407.3683} {\path{arXiv:1407.3683}},
  \href {http://dx.doi.org/10.1140/epjc/s10052-014-3149-z}
  {\path{doi:10.1140/epjc/s10052-014-3149-z}}.

\bibitem{Mohapatra+Senjanovic1981}
R.~N. Mohapatra, G.~Senjanovi\ifmmode~\acute{c}\else \'{c}\fi{},
  \href{http://link.aps.org/doi/10.1103/PhysRevD.23.165}{Neutrino masses and
  mixings in gauge models with spontaneous parity violation}, Phys. Rev. D 23
  (1981) 165--180.
\newblock \href {http://dx.doi.org/10.1103/PhysRevD.23.165}
  {\path{doi:10.1103/PhysRevD.23.165}}.
\newline\urlprefix\url{http://link.aps.org/doi/10.1103/PhysRevD.23.165}

\bibitem{Bilenky+Giunti2015}
S.~M. Bilenky, C.~Giunti, {Neutrinoless Double-Beta Decay: a Probe of Physics
  Beyond the Standard Model}, Int. J. Mod. Phys. A30~(04n05) (2015) 1530001.
\newblock \href {http://arxiv.org/abs/1411.4791} {\path{arXiv:1411.4791}},
  \href {http://dx.doi.org/10.1142/S0217751X1530001X}
  {\path{doi:10.1142/S0217751X1530001X}}.

\bibitem{PhysRevD.83.113003}
A.~Faessler, A.~Meroni, S.~T. Petcov, F.~\ifmmode~\check{S}\else
  \v{S}\fi{}imkovic, J.~Vergados,
  \href{http://link.aps.org/doi/10.1103/PhysRevD.83.113003}{Uncovering multiple
  $cp$-nonconserving mechanisms of
  $(\ensuremath{\beta}\ensuremath{\beta}{)}_{0\ensuremath{\nu}}$ decay}, Phys.
  Rev. D 83 (2011) 113003.
\newblock \href {http://dx.doi.org/10.1103/PhysRevD.83.113003}
  {\path{doi:10.1103/PhysRevD.83.113003}}.
\newline\urlprefix\url{http://link.aps.org/doi/10.1103/PhysRevD.83.113003}

\bibitem{Barry:2013xxa}
J.~Barry, W.~Rodejohann, {Lepton number and flavour violation in TeV-scale
  left-right symmetric theories with large left-right mixing}, JHEP 09 (2013)
  153.
\newblock \href {http://arxiv.org/abs/1303.6324} {\path{arXiv:1303.6324}},
  \href {http://dx.doi.org/{10.1007/JHEP09(2013)153}}
  {\path{doi:{10.1007/JHEP09(2013)153}}}.

\bibitem{SuhonenCivitarese1998}
J.~Suhonen, O.~Civitarese, Weak-interaction and nuclear-structure aspects of
  nuclear double beta decay, Phys. Rep. 300~(3-4) (1998) 123.
\newblock \href {http://dx.doi.org/10.1016/S0370-1573(97)00087-2}
  {\path{doi:10.1016/S0370-1573(97)00087-2}}.

\bibitem{PhysRevC.85.034316}
J.~Kotila, F.~Iachello,
  \href{http://link.aps.org/doi/10.1103/PhysRevC.85.034316}{Phase-space factors
  for double-$\ensuremath{\beta}$ decay}, Phys. Rev. C 85 (2012) 034316.
\newblock \href {http://dx.doi.org/10.1103/PhysRevC.85.034316}
  {\path{doi:10.1103/PhysRevC.85.034316}}.
\newline\urlprefix\url{http://link.aps.org/doi/10.1103/PhysRevC.85.034316}

\bibitem{StoicaMirea2013}
S.~Stoica, M.~Mirea, New calculations for phase space factors involved in
  double-beta decay, Phys. Rev. C 88~(3) (2013) 037303.
\newblock \href {http://dx.doi.org/10.1103/PhysRevC.88.037303}
  {\path{doi:10.1103/PhysRevC.88.037303}}.

\bibitem{Mirea2015}
M.~Mirea, T.~Pahomi, S.~Stoica, Values of the phase space factors involved in
  double beta decay, Rom. Rep. Phys. 67 (2015) 035503.

\bibitem{Stefanik+Dvornicky2015}
D.~\ifmmode~\check{S}\else \v{S}\fi{}tef\'anik, R.~Dvornick\'y,
  F.~\ifmmode~\check{S}\else \v{S}\fi{}imkovic, P.~Vogel,
  \href{http://link.aps.org/doi/10.1103/PhysRevC.92.055502}{Reexamining the
  light neutrino exchange mechanism of the
  $0\ensuremath{\nu}\ensuremath{\beta}\ensuremath{\beta}$ decay with left- and
  right-handed leptonic and hadronic currents}, Phys. Rev. C 92 (2015) 055502.
\newblock \href {http://dx.doi.org/10.1103/PhysRevC.92.055502}
  {\path{doi:10.1103/PhysRevC.92.055502}}.
\newline\urlprefix\url{http://link.aps.org/doi/10.1103/PhysRevC.92.055502}

\bibitem{Neacsu:2015uja}
A.~Neacsu, M.~Horoi, {An effective method to accurately calculate the phase
  space factors for $\beta^- \beta^-$ decay}, Adv. High Energy Phys. 2016
  (2016) 7486712.
\newblock \href {http://arxiv.org/abs/1510.00882} {\path{arXiv:1510.00882}},
  \href {http://dx.doi.org/10.1155/2016/7486712}
  {\path{doi:10.1155/2016/7486712}}.

\bibitem{Retamosa1995}
J.~Retamosa, E.~Caurier, F.~Nowacki, Neutrinoless double-beta decay of ca-48,
  Phys. Rev. C 51~(1) (1995) 371.
\newblock \href {http://dx.doi.org/10.1103/PhysRevC.51.371}
  {\path{doi:10.1103/PhysRevC.51.371}}.

\bibitem{HoroiStoicaBrown2007}
M.~Horoi, S.~Stoica, B.~A. Brown, Shell-model calculations of two-neutrino
  double-beta decay rates of $^{48}${Ca} with the {GXPF1A} interaction, Phys.
  Rev. C 75 (2007) 034303.
\newblock \href {http://dx.doi.org/10.1103/PhysRevC.75.034303}
  {\path{doi:10.1103/PhysRevC.75.034303}}.

\bibitem{HoroiStoica2010}
M.~Horoi, S.~Stoica, Shell model analysis of the neutrinoless double-beta decay
  of ca-48, Phys. Rev. C 81~(2) (2010) 024321.
\newblock \href {http://dx.doi.org/10.1103/PhysRevC.81.024321}
  {\path{doi:10.1103/PhysRevC.81.024321}}.

\bibitem{Horoi:2015gdv}
M.~Horoi, A.~Neacsu, {Analysis of mechanisms that could contribute to
  neutrinoless double-beta decay}, Phys. Rev. D93~(11) (2016) 113014.
\newblock \href {http://arxiv.org/abs/1511.00670} {\path{arXiv:1511.00670}},
  \href {http://dx.doi.org/10.1103/PhysRevD.93.113014}
  {\path{doi:10.1103/PhysRevD.93.113014}}.

\bibitem{HoroiBrown2013}
M.~Horoi, B.~A. Brown, Shell-model analysis of the xe-136 double beta decay
  nuclear matrix elements, Phys. Rev. Lett. 110~(22) (2013) 222502.
\newblock \href {http://dx.doi.org/10.1103/PhysRevLett.110.222502}
  {\path{doi:10.1103/PhysRevLett.110.222502}}.

\bibitem{Caurier2008}
E.~Caurier, J.~Menendez, F.~Nowacki, A.~Poves, Influence of pairing on the
  nuclear matrix elements of the neutrinoless beta beta decays, Phys. Rev.
  Lett. 100~(5) (2008) 052503.
\newblock \href {http://dx.doi.org/10.1103/PhysRevLett.100.052503}
  {\path{doi:10.1103/PhysRevLett.100.052503}}.

\bibitem{MenendezPovesCaurier2009}
J.~Menendez, A.~Poves, E.~Caurier, F.~Nowacki, Disassembling the nuclear matrix
  elements of the neutrinoless beta beta decay, Nucl. Phys. A 818~(3--4) (2009)
  139.
\newblock \href {http://dx.doi.org/10.1016/j.nuclphysa.2008.12.005}
  {\path{doi:10.1016/j.nuclphysa.2008.12.005}}.

\bibitem{Suhonen2011}
J.~Suhonen, On the double-beta decay of $^{70}$zn, $^{70}$zn, $^{70}$kr,
  $^{70}$zr, $^{70}$ru, $^{70}$pd, and $^{124}$sn, Nucl. Phys. A 864 (2011) 63.
\newblock \href {http://dx.doi.org/10.1016/j.nuclphysa.2011.06.021}
  {\path{doi:10.1016/j.nuclphysa.2011.06.021}}.

\bibitem{Suhonen2015}
J.~Hyvarinen, J.~Suhonen, Nuclear matrix elements for decays with light or
  heavy majorana-neutrino exchange, Phys. Rev. C 91 (2015) 024613.
\newblock \href {http://dx.doi.org/10.1103/PhysRevC.91.024613}
  {\path{doi:10.1103/PhysRevC.91.024613}}.

\bibitem{Simkovic2008}
F.~Simkovic, A.~Faessler, V.~Rodin, P.~Vogel, J.~Engel, Anatomy of the 0 nu
  beta beta nuclear matrix elements, Phys. Rev. C 77~(4) (2008) 045503.
\newblock \href {http://dx.doi.org/10.1103/PhysRevC.77.045503}
  {\path{doi:10.1103/PhysRevC.77.045503}}.

\bibitem{Simkovic2009}
F.~Simkovic, A.~Faessler, H.~Muether, V.~Rodin, M.~Stauf, 0 nu beta beta-decay
  nuclear matrix elements with self-consistent short-range correlations, Phys.
  Rev. C 79~(5) (2009) 055501.
\newblock \href {http://dx.doi.org/10.1103/PhysRevC.79.055501}
  {\path{doi:10.1103/PhysRevC.79.055501}}.

\bibitem{SimkovicRodin2013}
F.~Simkovic, V.~Rodin, A.~Faessler, P.~Vogel, $0\nu\beta\beta$ and
  $2\nu\beta\beta$ nuclear matrix elements, quasiparticle random-phase
  approximation, and isospin symmetry restoration, Phys. Rev. C 87 (2013)
  045501.
\newblock \href {http://dx.doi.org/10.1103/PhysRevC.87.045501}
  {\path{doi:10.1103/PhysRevC.87.045501}}.

\bibitem{Barea2013}
J.~Barea, J.~Kotila, F.~Iachello, Nuclear matrix elements for double-beta
  decay, Phys. Rev. C 87~(1) (2013) 014315.
\newblock \href {http://dx.doi.org/10.1103/PhysRevC.87.014315}
  {\path{doi:10.1103/PhysRevC.87.014315}}.

\bibitem{Barea2015}
J.~Barea, J.~Kotila, F.~Iachello, $0\nu\beta\beta$ and $2\nu\beta\beta$ nuclear
  matrix elements in the interacting boson model with isospin restoration,
  Phys. Rev. C 91 (2015) 034304.
\newblock \href {http://dx.doi.org/10.1103/PhysRevC.91.034304}
  {\path{doi:10.1103/PhysRevC.91.034304}}.

\bibitem{Rath2013}
P.~K. Rath, R.~Chandra, K.~Chaturvedi, P.~Lohani, P.~K. Raina, J.~G. Hirsch,
  Neutrinoless beta beta decay transition matrix elements within mechanisms
  involving light majorana neutrinos, classical majorons, and sterile
  neutrinos, Phys. Rev. C 88~(6) (2013) 064322.
\newblock \href {http://dx.doi.org/10.1103/PhysRevC.88.064322}
  {\path{doi:10.1103/PhysRevC.88.064322}}.

\bibitem{Rodriguez2010}
T.~R. Rodriguez, G.~Martinez-Pinedo, Energy density functional study of nuclear
  matrix elements for neutrinoless beta beta decay, Phys. Rev. Lett. 105~(25)
  (2010) 252503.
\newblock \href {http://dx.doi.org/10.1103/PhysRevLett.105.252503}
  {\path{doi:10.1103/PhysRevLett.105.252503}}.

\bibitem{Song2014}
L.~S. Song, J.~M. Yao, P.~Ring, J.~Meng, Relativistic description of nuclear
  matrix elements in neutrinoless double-beta decay, Phys. Rev. C 90~(5) (2014)
  054309.
\newblock \href {http://dx.doi.org/10.1103/PhysRevC.90.054309}
  {\path{doi:10.1103/PhysRevC.90.054309}}.

\bibitem{NeacsuHoroi2016}
M.~Horoi, A.~Neacsu, Shell model predictions for $^{124}$sn double-$\beta$
  decay, Phys. Rev. C 93 (2016) 024308.
\newblock \href {http://dx.doi.org/10.1103/PhysRevC.93.024308}
  {\path{doi:10.1103/PhysRevC.93.024308}}.

\bibitem{BUHRING1963472}
W.~B{\"u}hring,
  \href{http://www.sciencedirect.com/science/article/pii/0029558263902906}{Beta
  decay theory using exact electron radial wave functions}, Nuclear Physics 40
  (1963) 472 -- 488.
\newblock \href
  {http://dx.doi.org/http://dx.doi.org/10.1016/0029-5582(63)90290-6}
  {\path{doi:http://dx.doi.org/10.1016/0029-5582(63)90290-6}}.
\newline\urlprefix\url{http://www.sciencedirect.com/science/article/pii/0029558263902906}

\bibitem{HALPRIN1983335}
A.~Halprin, S.~Petcov, S.~Rosen,
  \href{http://www.sciencedirect.com/science/article/pii/0370269383912960}{Effects
  of light and heavy majorana neutrinos in neutrinoless double beta decay},
  Physics Letters B 125~(4) (1983) 335 -- 338.
\newblock \href
  {http://dx.doi.org/http://dx.doi.org/10.1016/0370-2693(83)91296-0}
  {\path{doi:http://dx.doi.org/10.1016/0370-2693(83)91296-0}}.
\newline\urlprefix\url{http://www.sciencedirect.com/science/article/pii/0370269383912960}

\bibitem{Towner+Hardy1995}
I.~Towner, J.~Hardy, Currents and their couplings in the weak sector of the
  standard model, Symmetries and fundamental interactions in nuclei (1995) 183.

\bibitem{Simkovic+Pantis1999}
F.~\ifmmode~\check{S}\else \v{S}\fi{}imkovic, G.~Pantis, J.~D. Vergados,
  A.~Faessler,
  \href{http://link.aps.org/doi/10.1103/PhysRevC.60.055502}{Additional nucleon
  current contributions to neutrinoless double $\ensuremath{\beta}$ decay},
  Phys. Rev. C 60 (1999) 055502.
\newblock \href {http://dx.doi.org/10.1103/PhysRevC.60.055502}
  {\path{doi:10.1103/PhysRevC.60.055502}}.
\newline\urlprefix\url{http://link.aps.org/doi/10.1103/PhysRevC.60.055502}

\bibitem{halzen1984quarks}
F.~Halzen, A.~Martin,
  \href{https://books.google.com/books?id=zwDvAAAAMAAJ}{Quarks and Leptons: An
  Introductory Course in Modern Particle Physics}, Wiley, 1984.
\newline\urlprefix\url{https://books.google.com/books?id=zwDvAAAAMAAJ}

\bibitem{behrens1982electron}
H.~Behrens, W.~B{\"u}hring,
  \href{https://books.google.com/books?id=b062AAAAIAAJ}{Electron Radial Wave
  Functions and Nuclear Beta-Decay}, The International Series of Monographs on
  Physics Series, Clarendon Press, 1982.
\newline\urlprefix\url{https://books.google.com/books?id=b062AAAAIAAJ}

\bibitem{PhysRevD.13.2567}
A.~Halprin, P.~Minkowski, H.~Primakoff, S.~P. Rosen,
  \href{http://link.aps.org/doi/10.1103/PhysRevD.13.2567}{Double-beta decay and
  a massive majorana neutrino}, Phys. Rev. D 13 (1976) 2567--2571.
\newblock \href {http://dx.doi.org/10.1103/PhysRevD.13.2567}
  {\path{doi:10.1103/PhysRevD.13.2567}}.
\newline\urlprefix\url{http://link.aps.org/doi/10.1103/PhysRevD.13.2567}

\bibitem{Deppisch2012}
F.~F. Deppisch, M.~Hirsch, H.~Pas, Neutrinoless double-beta decay and physics
  beyond the standard model, J. Phys. G 39~(12) (2012) 124007.
\newblock \href {http://dx.doi.org/10.1088/0954-3899/39/12/124007}
  {\path{doi:10.1088/0954-3899/39/12/124007}}.

\end{thebibliography}

\end{document}